\documentclass{article}
\begin{document}
\title{Interaction between a stationary electric charge and a stationary magnetic dipole}
\author{W.B.Bonnor}
\maketitle

\begin{abstract}
Using Einstein-Maxwell theory I investigate the gravitational field generated by
an electric charge and a magnetic dipole, both held in fixed positions, but
spinning with prescribed angular momenta.  There is a conical singularity
between them representing a strut balancing the gravitational attraction of their masses.
However, there is in general another singularity, which I call a torsion singularity.
I interpret this as a couple needed to maintain the spins at their prescribed
values.  It vanishes when the parameters obey a certain formula.

A conclusion of the work is that the charge and the magnet must spin relative to one another
unless constrained by a couple.
\end{abstract}

\begin{flushleft}
Comments: 11 pages, Journal reference: Class. Quantum Grav. 18 (2001) 2853.
\end{flushleft}

\section{Introduction}
In classical electromagnetism there is no interaction between an electric
charge and a magnetic dipole if they are at rest in the same Lorentz frame.
In general relativity the situation is different.  I showed [1] that a
charged magnetic dipole induces a rotational gravitational field which
drags inertial frames.

Let us now consider the gravitational field resulting from
a charge on the axis of a magnetic dipole, both at rest in a frame
Minkowskian at infinity.  Since both have mass, they need a strut between
them to counter their mutual gravitational attraction, and this is
represented by the usual well understood conical singularity.  However,
there is a second singularity between the charge and the dipole, of the
type previously noticed between uncharged spinning particles [2], and called
a torsion singularity.
This represents a couple stopping the charge and dipole from
spinning relative to one another.

To allow for spin \footnote{The spinning particles considered in this Letter are classical.}
it is useful to endow the charge and dipole with angular momenta.  To be precise,
let us consider two point masses, $m_{1}$ carrying electric charge $e$ and
angular momentum $h_{1}$, and $m_{2}$ with magnetic moment $M$ and angular momentum $h_{2}$.
Both particles are fixed in position on the $z$-axis (but allowed to spin
with the prescribed angular momenta), the first at $z=b>0$ and the second
at $z=-b$.  Applying Einstein-Maxwell theory and using a method of approximation
\footnote{For a different approach to the problem of charged magnetic spinning
particles see [6] [9] [10] [11].}
I find that there is a torsion singularity unless
\begin{equation}
2(m_{1}h_{2}+m_{2}h_{1})=eM.
\end{equation}
This means that if the parameters do not satisfy (1) there must be an applied
couple acting on the particles.

The plan of the Letter is as follows.  Section 2 contains the field equations
and the metric of spacetime.  In section 3 I derive the first approximation
to the solution, and give a partial solution of the second approximation.
This enables me in section 4 to exhibit the torsion singularity and to
derive formula (1).

\section{Field equations and metric}
The field equations are those of Einstein-Maxwell theory for empty space:
\begin{eqnarray}
R^{i}_{k}&=&2F^{ia}F_{ka} -\frac{1}{2}\delta^{i}_{k}F^{ab}F_{ab},\\
F_{ik}&=&A_{i,k}-A_{k,i},\\
F^{ik}_{;k}&=&0.
\end{eqnarray}
Here $R^{i}_{k}$ is the Ricci tensor, $F_{ik}$ the electromagnetic field
tensor and $A_{i}$ the vector potential; a comma denotes partial differentiation
and a semi-colon covariant differentiation.

We use the stationary axially symmetric metric
\begin{equation}
ds^{2}=-f^{-1}e^{\nu}(dz^{2}+dr^{2})-ld\theta^{2}-2nd\theta dt+fdt^{2},
.\end{equation}
where $f, \nu, l, n$ are functions of $z$ and $r$ only.  The coordinates will be
numbered
\[ x^{1}=z, x^{2}=r, x^{3}=\theta, x^{4}=t.\]
The components of the vector potential are
\[A_{i}=(0,0,\psi,\phi)\]
and these too are functions of $z$ and $r$ only.

Let
\[ \Delta^{2}=lf+n^{2}; \]
then we find that the field equations (2) and (3) imply
\[ R^{3}_{3}+R^{4}_{4}=-\Delta^{-1}fe^{-\nu}(\Delta_{11}+\Delta_{22})=0, \]
where suffices $1$ and $2$ mean differentiation with respect to $z$ and $r$.
Assuming $\Delta$ to be a monotonically increasing function of $r$, we may,
without loss of generality [3], take the solution of this equation
to be
\[ \Delta = r\]
so that
\begin{equation}
lf+n^{2}=r^{2}.
\end{equation}

With $\Delta$ chosen in this way the field equations (2) - (4) can be
written as follows:
\begin{eqnarray}
R_{11}+R_{22}&=&\nu_{11}+\nu_{22}-f^{-1}\nabla^{2}f+\frac{3}{2}f^{-2}(f_{1}^{2}+f_{2}^{2})\nonumber \\
             & &-\frac{1}{2}r^{-2}f^{2}(w_{1}^{2}+w_{2}^{2})=0,\\
R_{11}-R_{22}&=&r^{-1}\nu_{2}+\frac{1}{2}f^{-2}(f_{1}^{2}-f_{2}^{2})+\frac{1}{2}r^{-2}f^{2}(w_{2}^{2}-w_{1}^{2})\nonumber\\
             &=&2r^{-2}[f(\psi_{2}^{2}-\psi_{1}^{2})+l(\phi_{1}^{2}-\phi_{2}^{2})+2n(\phi_{2}\psi_{2}-\phi_{1}\psi_{1})],\\
2R_{12}&=&-r^{-1}\nu_{1}+f^{-2}f_{1}f_{2}-r^{-2}f^{2}w_{1}w_{2}\nonumber\\
       &=&4r^{-2}[l\phi_{1}\phi_{2}-f\psi_{1}\psi_{2}-n(\phi_{1}\psi_{2}+\phi_{2}\psi_{1})],\\
R^{4}_{4}-R^{3}_{3}-2wR^{3}_{4}&=&e^{-\nu}[-\nabla^{2}f+f^{-1}(f_{1}^{2}+f_{2}^{2})-r^{-2}f^{3}(w_{1}^{2}+w_{2}^{2})]\nonumber\\
                              &=&-2r^{-2}e^{-\nu}[(r^{2}+n^{2})(\phi_{1}^{2}+\phi_{2}^{2})+2fn(\phi_{1}\psi_{1}+\phi_{2}\psi_{2})\nonumber \\
                              & &+f^{2}(\psi_{1}^{2}+\psi_{2}^{2})],\\
2R^{3}_{4}&=&-r^{-2}f^{2}e^{-\nu}[f\nabla^{*2}w+2(f_{1}w_{1}+f_{2}w_{2})]\nonumber\\
          &=&4r^{-2}fe^{-\nu}[n(\phi_{1}^{2}+\phi_{2}^{2})+f(\phi_{1}\psi_{1}+\phi_{2}\psi_{2})],\\
r^{2}\nabla^{2}\phi&=&\sum_{i=1}^{2}[\phi_{i}(lf_{i}+nn_{i})-\psi_{i}(nf_{i}-fn_{i})],\\
r^{2}\nabla^{2}\psi&=&\sum_{i=1}^{2}[\psi_{i}(fl_{i}+nn_{i})-\phi_{i}(ln_{i}-nl_{i})],
\end{eqnarray}
where I have put
\begin{eqnarray}
w&=&nf^{-1},\\
\nabla^{2}X&=&X_{11}+X_{22}+r^{-1}X_{2},\nonumber \\
\nabla^{*2}X&=&X_{11}+X_{22}-r^{-1}X_{2},\nonumber
\end{eqnarray}

The structure of eqns (7)-(13) is as follows: (10)-(13) give $f, w, \phi, \psi$ in terms of their sources,and $\nu$ is determined
by (7)-(9), in which there is redundancy.

\section{Approximate solution}
We seek a solution for the two point masses described in the
Introduction.  As the solution is to be stationary the particles must be
subject to constraints, and these will be represented in our solution by
singularities.

I assume that the solution can be expanded in powers and products of the six
parameters $m_{1}, m_{2}, e, M, h_{1}, h_{2}$, and use a method of approximation.
For the zeroth approximation I take Minkowski spacetime with no electromagnetic
field, so that in (4)
\begin{equation}
\stackrel{(0)}{\nu}=0, \stackrel{(0)}{f}=1,  \stackrel{(0)}{l}=r^{2}, \stackrel{(0)}{n}=0.
\end{equation}
The first (or linear) approximation arises when one ignores all non-linear
terms in (6)-(14), whence one obtains
\begin{equation}
\stackrel{(1)}{\nu}=\stackrel{(1)}{C}, \stackrel{(1)}{l}=-r^{2}\stackrel{(1)}{f},
\nabla^{2}\stackrel{(1)}{f}=0, \nabla^{*2}\stackrel{(1)}{w}=0, \stackrel{(1)}{n}=\stackrel{(1)}{w},
\nabla^{2}\stackrel{(1)}{\phi}=0, \nabla^{*2}\stackrel{(1)}{\psi}=0,
\end{equation}
where $\stackrel{(1)}{C}$is an arbitrary constant which we henceforth put equal
to zero because a non-zero value would imply the existence of a cosmic string.

We now choose solutions of (16) representing the sources specified in section 1:
\begin{eqnarray}
\stackrel{(1)}{f}&=&-\sum_{i=1}^{2}\frac{2m_{i}}{R_{i}},\\ \stackrel{(1)}{w}&=&\sum_{i=1}^{2}\frac{2h_{i}r^{2}}{R_{i}^{3}},\\
\stackrel{(1)}{\phi}&=&\frac{e}{R_{1}},\\ \stackrel{(1)}{\psi}&=&\frac{Mr^{2}}{R_{2}^{3}}
\end{eqnarray}
where $R_{1}=\mid [r^{2}+(z-b)^{2}]^{1/2} \mid, R_{2}=\mid [r^{2}+(z+b)]^{1/2} \mid$
are both greater than the event horizon radii of the particles.  The factor 2 is inserted into (18) to ensure agreement with the
linear approximation to the Kerr solution.

Proceeding to the second approximation we insert (17-(20) into (7)-(13),
ignoring all terms of the third and higher orders.  This gives four
equations of Poisson type for the second approximation to the essential
unknowns, denoted $\stackrel{(2)}{f}, \stackrel{(2)}{w}, \stackrel{(2)}{\phi},
\stackrel{(2)}{\psi}; \stackrel{(2)}{\nu}$ is then obtainable from (7)-(9), and
$\stackrel{(2)}{n}$ and $\stackrel{(2)}{l}$ from (14) and (6).

The second approximation can be completely solved after long calculations,
but here I solve it only for the function $\stackrel{(2)}{w},$
since this gives rise to the torsion singularity representing the rotational
constraint which is my main interest in this Letter.

The second approximation to (11) may be written
\begin{equation}
\nabla^{*2}\stackrel{(2)}{w}=-2(\stackrel{(1)}{f}_{1}\stackrel{(1)}{w}_{1}+\stackrel{(1)}{f}_{2}\stackrel{(1)}{w}_{2})
-4(\stackrel{(1)}{\phi}_{1}\stackrel{(1)}{\psi}_{1}+\stackrel{(1)}{\phi}_{2}\stackrel{(1)}{\psi}_{2}).
\end{equation}
All the terms on the right are specified in (17)-(20), and inserting these
we find that (21) becomes
\begin{eqnarray*}
\nabla^{*2}\stackrel{(2)}{w}&=&\frac{4(2m_{1}h_{2}-eM)r^{2}}{R_{1}^{3}R_{2}^{5}}[r^{2}+(z-5b)(z+b)]\\
                            & &+\frac{8m_{2}h_{1}r^{2}}{R_{1}^{5}R_{2}^{3}}[r^{2}+(z+5b)(z-b)] +8\sum_{i=1}^{2}\frac{m_{i}h_{i}r^{2}}{R_{i}^{6}}.
\end{eqnarray*}
A calculation then shows that the solution is
\begin{eqnarray}
\stackrel{(2)}{w}&=&\frac{(2m_{1}h_{2}-eM)}{2b^{2}R_{1}R_{2}^{3}}[r^{4}+2r^{2}(z^{2}+bz+2b^{2})+(z-b)(z+b)^{3}]\nonumber\\
                 & &+\frac{m_{2}h_{1}}{b^{2}R_{1}^{3}R_{2}}[r^{4}+2r^{2}(z^{2}-bz+2b^{2})+(z+b)(z-b)^{3}]\nonumber \\
                 & &+2\sum_{i=1}^{2}\frac{m_{i}h_{i}r^{2}}{R_{i}^{4}}+\stackrel{(2)}{K},
\end{eqnarray}
where $\stackrel{(2)}{K}$ is an arbitrary constant.

(22) is a particular integral of (21): I insert no complementary
function because I assume that the sources have already been introduced in
the first approximation (17)-(20).  We shall also need $n$: from (14)
this is given by
\begin{equation}
\stackrel{(1)}{n}=\stackrel{(1)}{w}; \stackrel{(2)}{n}=\stackrel{(2)}{w}+\stackrel{(1)}{f}\stackrel{(1)}{w}.
\end{equation}

\section{Interpretation}
One condition for regularity on the symmetry axis is that $\nu=0$ there. This
is well known in the static non-electromagnetic case [3], in which it cannot
be satisfied: there is a conical singularity on the axis representing a strut (or strings)
supporting the particles against their mutual
gravitational attraction.  A similar singularity is, in general, present in
our system, though it is modified by the presence of the spins and electromagnetic
charges.  I shall not consider this singularity in this Letter.

Another condition for regularity on the symmetry axis is that $n=0$ there:
for otherwise, from (6), $l$ must change sign.  If this happens there exist closed timelike curves,
$r=0$ is no longer a spatial axis of symmetry, and the character of spacetime is
radically different.  Following Letelier [4] [5] I refer to a part of the axis
on which $n\neq 0$ as a torsion singularity.  In our case we see from (18)
and (23) that $n=0$ on $r=0$ in the first approximation, but in the second
approximation $\stackrel{(2)}{n}=0$ requires $\stackrel{(2)}{w}=0$ there.  Using
$\stackrel{(2)}{K}$ we can make $\stackrel{(2)}{w}$ vanish on the axis
{\em either} for $\mid z \mid<b$ {\em or} for $\mid z \mid >b$, but not for
both unless
\[eM=2(m_{1}h_{2}+m_{2}h_{1}).\;\;\;\;\;\;\;\;\;\;\;\;\;\;(1) \]

{\em There will be no torsion singularity if (1) is satisfied.}  Otherwise
there must be a singularity which, I have argued previously [2]
for a configuration of uncharged spinning particles, represents a couple
maintaining the spins at values which do not satisfy (1).
A remarkable feature of (1) is that the distance between
the particles does not occur in it.

In the special case in which the particles have no angular momentum (i.e. when
$h_{1}=h_{2}=0$), there will be a singularity representing a couple needed
to keep them from spinning.  It is tempting to conjecture that, if the
two particles were placed in position, supported against collapse, but allowed to spin,
their angular momenta would change until (1) was satisfied, and that the system would then remain in a steady state.

If we introduce the angular momenta per unit mass $a_{1}, a_{2}$, (1)
may be written
\[a_{1}+a_{2}=eM/2m_{1}m_{2}, \]
so that if either $e$ or $M$ is zero,
\[a_{1}+a_{2}=0,\]
a relation which has appeared several times in a similar context [2] [6] [7] [8].

Finally it should be remembered that the results of this Letter have been
derived from the first and second approximations to the equations (7)-(13).
Work with higher approximations may affect the conclusions, in particular
formula (1).

\section*{References}
{[1]} W.B.Bonnor, Phys. Lett. A158 (1991) 23.\\
{[2]} W.B.Bonnor, Class. Quantum Grav. 18 (2001) 1381.\\
{[3]} J.L.Synge, Relativity:The General Theory (North-Holland,Amsterdam, 1960) p.309.\\
{[4]} P.S.Letelier, Class. Quantum Grav. 12 (1995) 471.\\
{[5]} P.S.Letelier and S.R.de Oliveira, Phys. Lett. A238 (1998) 101.\\
{[6]} W.B.Bonnor and J.P.Ward, Commun. math. Phys. 28 (1972) 323.\\
{[7]} T.J.T.Spanos, Phys. Rev. D9 (1974) 1633.\\
{[8]} N.Bret\'{o}n and V.S.Manko, Class. Quantum Grav. 12 (1995) 1969.\\
{[9]} N.Bret\'{o}n, V.S.Manko and J.A.S\'{a}nchez, Class. Quantum Grav. 16 (1999) 3725.\\
{[10]}V.S.Manko, J.D.Sanabria-G\'{o}mez and O.V.Manko, Phys. Rev. 62 (2000) 044048.\\
{[11]}V.S.Manko, E.Ruiz and J.D.Sanabria-G\'{o}mez, Class. Quantum Grav. 17 (2000) 3881.
\end{document}